\documentclass[a4paper,fleqn,usenatbib]{mnras}
\usepackage{txfonts,times}
\usepackage[T1]{fontenc}
\usepackage{ae,aecompl}
\usepackage{graphicx}	

\title[Galaxy segregation inside filaments at $z\simeq 0.7$]{The VIMOS Public Extragalactic Redshift Survey (VIPERS): Galaxy segregation inside filaments at $z\simeq 0.7$}
\author[Malavasi et al.]{ 
N. Malavasi$^{1,2,}$\thanks{E-mail: \href{mailto: nicola.malavasi@lam.fr}{nicola.malavasi@lam.fr} / \href{mailto: nicola.malavasi@unibo.it}{nicola.malavasi@unibo.it}},
S. Arnouts$^{2}$, 
D. Vibert$^{2}$,
S. de la Torre$^{2}$,
T. Moutard$^{2}$,
C. Pichon$^{3,4}$,
\newauthor 
I. Davidzon$^{2}$,
K.~Kraljic$^{2}$,
M.~Bolzonella$^{5}$,
L.~Guzzo$^{6,8}$,   
B.~Garilli$^{7}$,
M.~Scodeggio$^{7}$,
\newauthor
B.~R.~Granett$^{6}$,
U.~Abbas$^{9}$,
C.~Adami$^{2}$,
D.~Bottini$^{7}$,
A.~Cappi$^{5}$,
O.~Cucciati$^{1,5}$,
\newauthor
P.~Franzetti$^{7}$,
A.~Fritz$^{7}$,
A.~Iovino$^{6}$, 
J.~Krywult$^{10}$,
V.~Le Brun$^{2}$,
O.~Le F\`evre$^{2}$,
\newauthor 
D.~Maccagni$^{7}$,
K.~Ma{\l}ek$^{11}$,
F.~Marulli$^{1,5,12}$,
M.~Polletta$^{7}$,
A.~Pollo$^{11,13}$,
L.~Tasca$^{2}$,
\newauthor 
R.~Tojeiro$^{14}$,
D.~Vergani$^{5,15}$,
A.~Zanichelli$^{16}$,
J.~Bel$^{17}$,
E.~Branchini$^{18}$,  
J.~Coupon$^{19}$, 
\newauthor 
G.~De Lucia$^{20}$, 
Y.~Dubois$^{3}$,
A.~Hawken$^{6}$,
O.~Ilbert$^{2}$,
C.~Laigle$^{3}$,
L.~Moscardini$^{1,5,12}$,
\newauthor 
T. ~Sousbie$^{3}$,
M.~Treyer$^{2}$  
and G.~Zamorani$^{5}$
\\
$^{1}$Universit\`{a} di Bologna -- Dipartimento di Fisica e Astronomia (DIFA), v.le Berti Pichat 6/2 - 40127 Bologna, Italia\\  
$^{2}$Aix Marseille Universit\'e, CNRS, Laboratoire d'Astrophysique de Marseille, UMR 7326, 13388, Marseille, France\\  
$^{3}$Institute d'Astrophysique de Paris, UMR7095 CNRS, Universit\'{e} Pierre et Marie Curie, 98 bis Boulevard Arago, 75014 Paris, France\\ 
$^{4}$Korea Institute of Advanced Studies (KIAS) 85 Hoegiro, Dongdaemun-gu, Seoul, 02455, Republic of Korea\\   
$^{5}$INAF - Osservatorio Astronomico di Bologna, via Ranzani 1, I-40127, Bologna, Italy\\ 
$^{6}$INAF - Osservatorio Astronomico di Brera, Via Brera 28, 20122 Milano, via E. Bianchi 46, 23807 Merate, Italy\\ 
$^{7}$INAF - IASF Milano, via Bassini 15, 20133 Milano, Italy\\ 
$^{8}$Universit\`{a} degli Studi di Milano, via G. Celoria 16, 20130 Milano, Italy\\ 
$^{9}$INAF - Osservatorio Astronomico di Torino, 10025 Pino Torinese, Italy\\ 
$^{10}$Institute of Physics, Jan Kochanowski University, ul. Swietokrzyska 15, 25-406 Kielce, Poland\\ 
$^{11}$National Centre for Nuclear Research, ul. Hoza 69, 00-681 Warszawa, Poland\\ 
$^{12}$INFN, Sezione di Bologna, viale Berti Pichat 6/2, I-40127 Bologna, Italy\\ 
$^{13}$Astronomical Observatory of the Jagiellonian University, Orla 171, 30-001 Cracow, Poland\\ 
$^{14}$School of Physics and Astronomy, University of St Andrews, St Andrews KY16 9SS, UK \\ 
$^{15}$INAF - IASF Bologna, via Gobetti 101, I-40129 Bologna, Italy\\ 
$^{16}$INAF - Istituto di Radioastronomia, via Gobetti 101, I-40129, Bologna, Italy\\ 
$^{17}$Aix Marseille Universit\'e, CNRS, CPT, UMR 7332, 13288 Marseille, France\\   
$^{18}$Dipartimento di Matematica e Fisica, Universit\`{a} degli Studi Roma Tre, via della Vasca Navale 84, 00146 Roma, Italy\\ 
$^{19}$Astronomical Observatory of the University of Geneva, ch. d'Ecogia  16, 1290 Versoix, Switzerland\\ 
$^{20}$INAF - Osservatorio Astronomico di Trieste, via G. B. Tiepolo 11, 34143 Trieste, Italy\\}
           
\begin{document}

\date{Accepted 2016 November 02. Received 2016 November 02; in original form 2016 September 14}

\pagerange{\pageref{firstpage}--\pageref{lastpage}} \pubyear{2016}

\maketitle

\label{firstpage}

\begin{abstract}
We present the first quantitative detection of large-scale filamentary structure at $z \simeq 0.7$ in the large cosmological volume probed by the VIMOS Public Extragalactic Redshift Survey (VIPERS). We use simulations to show the capability of VIPERS to recover robust topological features in the galaxy distribution, in particular the filamentary network. We then investigate how galaxies with different stellar masses and stellar activities are distributed around the filaments and find a significant
segregation, with the most massive or quiescent galaxies being closer to the filament axis than less massive or active galaxies. The signal persists even after down-weighting the contribution of peak regions. Our results suggest that massive and quiescent galaxies assemble their stellar mass through successive mergers during their migration along filaments towards the nodes of the cosmic web. On the other hand, low-mass star-forming galaxies prefer the outer edge of filaments,  
a vorticity rich region dominated by smooth accretion, as predicted by the recent spin alignment theory. This emphasizes the role of large scale cosmic flows in shaping galaxy properties.
%
%
\end{abstract}

\begin{keywords}
Cosmology: observations -- Cosmology: large-scale structure of Universe -- Galaxies: evolution -- Galaxies: high-redshift -- Galaxies: statistics.
\end{keywords}

\section{Introduction}
\label{intro}
A major success of the $\Lambda$CDM model is its ability to reproduce the wealth of large-scale structures (LSS) observed in the galaxy distribution \citep[e.g.][]{Springel2006}. These structures arise from the growth of primordial, nearly Gaussian, matter density fluctuations under the effect of gravity. In this process, matter departs from underdense regions and flows through dense sheets that wind up, forming filaments along which matter drifts and progressively gets accreted onto high-density peaks. This leads to a cosmic web (CW) where dense nodes are connected by filaments, framing walls separated by large voids \citep{Bond1996}. The baryonic gas follows the gravitational potential gradients imposed by the dark matter distribution, then is shocked, forming (among other structures) tenuous ionised hydrogen filaments, the intergalactic medium (IGM), in
which galaxies can form. These filaments are regions where gas, momentum, and energy are exchanged through the complex processes of infall and outflow. While it has long been established that the local density environment on typical scales below a few Mpc plays a role in shaping galaxy properties \citep[see \textit{e.g.}][]{Dressler1980, zehavi2005}, the extent to which large-scale anisotropic structures and the tidal field of the CW influence the evolution of galaxies (and subsequently properties such as morphology, accretion mode, and merging rate) still remains an open issue.

There is significant numerical evidence that large-scale environment has an impact on the formation and evolution of galaxies. In particular, N-body dark matter simulations have shown that the spin and the shape of dark matter haloes depend on the
large-scale environment in which they reside \citep{AragonCalvo2007a,Hahn2007, Sousbie2009}. Moreover, using hydrodynamical simulations, \citet{Keres2005} found that at high redshift cold streams can penetrate deep inside haloes and feed galaxies with fresh gas to sustain intense star formation (SF) activity \citep[see also][]{Dekel2009,DekelBirnboim2006}. \citet{Pichon2011} proposed that the filamentary flows advect angular momentum onto the disks of galaxies and that the spin of newly formed galaxies tends preferentially to be parallel to the axis of their closest filament. \citet{Codis2012} quantified a mass transition, with the most massive haloes ending up with a spin perpendicular to the filaments as a result of successive mergers along the filaments. These results have been extended to galaxies by \citet{Dubois2014} with the state-of-the-art hydrodynamical simulation Horizon-AGN \citep[see also][for a theoretical motivation for this transition based on constrained tidal torque theory]{codis2015}. On the side of observations, the correlation between the spin of galaxies and their filaments or sheets has been recently detected in the Sloan Digital Sky Survey \citep{Tempel2013, Zhang2013,Trujillo2006}. This remarkable result confirms the role played by the large scale dynamical environment in the evolution of galaxies\footnote{This intrinsic alignment is also a source of systematics affecting the cosmic shear signal \cite[e.g.][]{Chisari2015}.}, which is usually neglected in galaxy formation models \citep[\textit{e.g.}][]{zehavi2005,guo2011}.

In the local Universe ($z \le 0.3$) the large galaxy redshift surveys 2dF \citep{Colless2001}, SDSS \citep{York2000}, and GAMA
\citep{Driver2011} have captured the CW in great detail and led to several analyses showing variations of individual \citep{Beygu2013,Alpaslan2015,Alpaslan2016} and statistical \citep{Eardley2015, Martinez2016} galaxy properties as a function of their CW environment. At higher redshifts, because of the small volumes probed and the low sampling rates achieved by redshift surveys, this kind of analysis has not been possible until recently. The state-of-the-art redshift survey VIPERS \citep{Guzzo2014} overcomes these limitations by probing a volume equivalent to the local 2dF survey with a dense spectroscopic sampling of galaxies. It allowed the first measurement of the growth rate of the LSS at $0.5 \le z \le 1.2$ \citep{delaTorre2013} and the fine mapping of the CW at an epoch when the Universe was about half its current age.

In this paper, we exploit the final, complete sample of the VIPERS survey to detect the filamentary structure of the CW at high redshift and to study the correlation between galaxy properties and their distance to the closest filament. After describing the data in Sect. \ref{data}, we illustrate in Sect. \ref{method} the filament reconstruction with the Discrete Persistent Structure Extractor \citep[DisPerSE,][]{Sousbie2011a} on mock samples of the VIPERS survey and its fidelity. We present the application to the VIPERS data in Sect. \ref{results}, where we also report a significative mass and type segregation within filaments. Finally, we discuss in Sect. \ref{conclusions} our results within the current paradigm of galaxy assembly. Unless stated otherwise, we assume the \citet{PlanckXIII} cosmology with $H_0 = 67.51$ km s$^{-1}$ Mpc$^{-1}$, $h = H_0 / 100$, $\Omega_m = 0.3121$, and $\Omega_{\Lambda} = 0.6879$.

\section{Data}
\label{data}
The VIMOS Public Extragalactic Redshift Survey (VIPERS) is a magnitude-limited spectroscopic galaxy survey to $i_{AB} \le 22.5$. It covers an overall footprint of about $16 \deg^2$ and $8 \deg^2$ in the W1 and W4 fields of the CFHTLS-Wide imaging survey, respectively. VIPERS spectra were collected in low resolution mode, $R = 230$, leading to a radial velocity error of $\sigma_v = 175 (1+z_{\rm spec})$ km s$^{-1}$. The spectroscopic targets were pre-selected in a colour-colour space to remove galaxies below $z=0.5$, which coupled with an optimized observing strategy, provides an average effective sampling rate of about $40\%$. We refer the reader to the survey description papers by \citet{Guzzo2014} and \citet{Garilli2014} and to the parallel paper by \citet{Scodeggio2016} for more details.

In this work we use the final galaxy sample, described in the latter paper (the so-called PDR-2). We consider only the most
secure redshifts, corresponding to quality flag $\ge 2$ in the VIPERS grading scheme (confidence level, CL $> 97 \%$). The mean number density of galaxies, $\overline{n}(z)$, varies significantly at the redshift boundaries of the survey, due to the
magnitude limit, the target sampling rate, and the colour selection \citep[see][for details]{delaTorre2013,Guzzo2014}. For this reason we limit this analysis to the 50\,980 galaxies in the range $0.5\le z\le 0.85$, where the typical spatial resolution in terms of mean inter-galaxy separation, $\langle D_z\rangle \sim \overline{n}{(z)}^{-1/3}$, is the highest ($7.7 < \langle D_z \rangle/\mathrm{Mpc} < 10$). These values are comparable with those of the GAMA survey \citep[][ with $4.6 < \langle D_z \rangle/\mathrm{Mpc} < 8.8$ for $0.1 < z < 0.3$ and $r \le 19.8$]{Driver2011}, and make VIPERS the first galaxy redshift survey well suited for studying the CW at high redshift. The stellar masses for the objects in our sample and the classification between active and passive populations were derived according to \citet[][]{Moutard2016b}, on the basis of the SED-fitting analysis of the multi-wavelength data collected in the VIPERS regions\footnote{The VIPERS-MLS survey: \href{http://cesam.lam.fr/vipers-mls/}{http://cesam.lam.fr/vipers-mls/}} \citep[][]{Moutard2016a}.

\section{Detecting filaments in the VIPERS survey }
\label{method}
\paragraph*{The DisPerSE code.}
In order to trace the CW in VIPERS, we rely on the Discrete Persistent Structure Extractor \citep[DisPerSE, see][for a complete description]{Sousbie2011a,Sousbie2011b}. DisPerSE identifies filaments as ridges in the density field (calculated using the Delaunay Tessellation Field Estimator, DTFE). DisPerSE uses the discrete Morse theory to extract critical points, where the gradient of the density field vanishes (e.g. maxima and saddle points), and the field lines connecting them. It then pairs the critical points in topological features, called \textquotedblleft critical pairs\textquotedblright, using the persistent homology theory. The robustness of each feature (including the filaments) is assessed by the relative density contrast of its critical pair, the so-called persistence, which is chosen to pass a certain signal-to-noise (S/N) threshold. The noise level is defined relative to the variance of persistence values obtained from random sets of points. Because DisPerSE is based on a topologically-motivated algorithm, it is both very robust and flexible through the choice of the persistence threshold. Since it filters out the sampling noise, it enables an unsmoothed density field, more noisy but less biased, to be analysed. By construction it is also multi-scale: it builds a network which adapts naturally to the uneven sampling of observed catalogues. The persistence threshold is calibrated on mocks to account for the specific design of VIPERS. To prevent the spurious detection of filaments near the edges of the survey, DisPerSE encloses each field in a larger volume. New particles are added by interpolating the density field measured at the boundary of the survey \citep[see \textit{e.g.}][]{Sousbie2011b}.

\paragraph*{Tests on VIPERS mock galaxy catalogues.}
\label{mocktests}
We test the performance of DisPerSE on an updated version of the VIPERS mock galaxy catalogues described in \citet[][]{delaTorre2013} matching the VIPERS final geometry. The upgrades are described in the parallel paper by \citet{delaTorre2016}. The parent catalogues include all the galaxies down to the magnitude limit $i_{AB}=22.5$ together with the correct selection function at $0.4<z<0.6$ due to the VIPERS colour pre-selection. The VIPERS-like catalogues are built from the parent ones with all the observational effects  applied (\textit{i.e.} pointing strategy, target sampling rate, gaps between VIMOS quadrants, photometric mask and random errors on redshift).

The impact of observational biases on the skeleton reconstruction can be assessed by comparing the skeletons obtained from the parent and the VIPERS-like mock catalogues. To quantify the differences between two skeletons we define a pseudo-distance between the two skeletons to be compared. In practice, a skeleton {\it S$_a$} is composed by $N_a$ short straight segments, {\it s$_a^i$}.  We define the pseudo-distance from a skeleton {\it   S$_a$} to a skeleton {\it S$_b$}, {\it D(S$_a$, S$_b$)}, as the probability distribution function (PDF) of the distances between each segment of {\it S$_a$}, {\it s$_a^i$}, and its closest segment in {\it S$_b$}, {\it s$_b^j$} \citep[][]{Sousbie2009}. The distributions {\it D(S$_a$, S$_b$)} and {\it D(S$_b$, S$_a$)} are composed by $N_a$ and $N_b$ distances respectively. There is no reason for the pseudo-distance {\it D(S$_a$,   S$_b$)} to be identical to {\it D(S$_b$, S$_a$)}. Indeed the discrepancy between the two PDFs is related to the differences between the two skeletons\footnote{Note that by construction the Delaunay tessellation provides a simple way to reconnect LSS features with smoothed variations in density such as filaments across large gaps \citep[][]{AragonCalvo2015}. For this reason and based on the results on simulations discussed here, we do not apply any correction for the gaps in the VIPERS survey.}.

\begin{figure}
\centering
\includegraphics[width=0.5\textwidth]{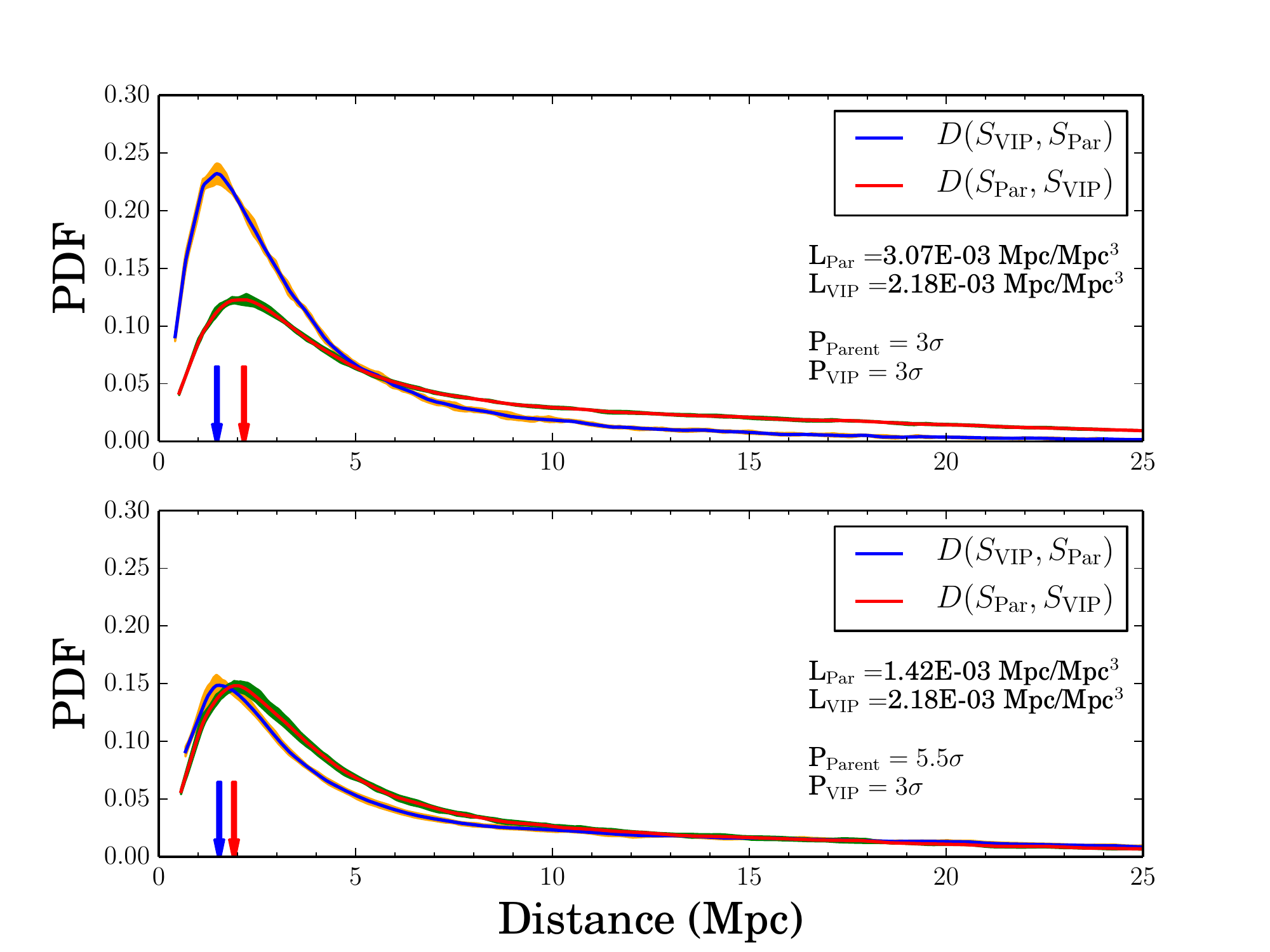}
\caption{PDFs of the pseudo-distances (defined in Sect. \ref{method}) between the Parent and VIPERS-like skeletons. Blue lines and orange shaded areas refer to {\it S$_{\rm VIP}$} projected onto {\it S$_{\rm Par}$} while red lines and green shaded areas are the reverse. Solid lines correspond to the mean of 10 mocks and the shaded areas enclose the $1\sigma$ r.m.s. (please note that uncertainties are negligible). Arrows  show the modes of the distributions. \textit{Top}: The two skeletons are extracted with a $3\sigma$ persistence threshold. 
\textit{Bottom}: {\it S$_{\rm Par}$} is extracted with a $5.5\sigma$ threshold.} 
\label{parent_vs_spectroscopic}
\end{figure}

Fig. \ref{parent_vs_spectroscopic} presents the PDFs of the pseudo-distances obtained by comparing the skeletons {\it S$_{\rm Par}$} and {\it S$_{\rm VIP}$} measured from the Parent and VIPERS-like catalogues, respectively, in the redshift range $0.5 < z < 0.85$. In the upper panel, DisPerSE is run with a persistence threshold of $3\sigma$ in both catalogues. This threshold guarantees that less than 1\% of critical pairs are spurious, as tested on random field simulations \citep{Sousbie2011a}. {\it S$_{\rm VIP}$} is reconstructed with less accuracy and detail due to the lower sampling. An estimate of the uncertainty in the location of the filaments is given by the modes of the PDFs which do not peak at distance $D\sim 0$ (corresponding to a perfect match between the segments of the two skeletons) but  $D\sim 1.5-2$ Mpc (marked as vertical arrows). The asymmetry between {\it D($S_{\rm Par}, S_{\rm VIP}$)} and {\it D($S_{\rm VIP}, S_{\rm Par}$)} reflects the fact that {\it S$_{\rm Par}$} (full sampling) has much more details (\textit{i.e.} the leaves of branches, or small branches) which have no counterpart in $S_{\rm VIP}$.  On the other hand  90\% of the segments of $S_{\rm VIP}$ have a counterpart in $S_{\rm Par}$ with distances $D\le 10$ Mpc, illustrating the small fraction of spurious filaments. We also report the average length of the VIPERS-like ({\it L$_{\rm VIP}$}) and the Parent ({\it L$_{\rm Par}$}) skeletons, defined as the total skeleton length divided by the survey volume (expressed in Mpc/Mpc$^3$). {\it L$_{\rm VIP}$} is shorter than {\it L$_{\rm Par}$} as expected for a skeleton with fewer details. In the lower panel, the persistence threshold on the Parent catalogue is increased to 5.5$\sigma$. The length, {\it L$_{\rm Par}$}, is  shortened with only the most significant filaments still present. The two PDFs for {\it D($S_{\rm VIP}, S_{\rm Par}$)} and {\it D($S_{\rm Par}, S_{\rm VIP}$)} are now much closer in amplitude and shape. Most of the segments in $S_{\rm Par}$ (75\%) have a counterpart in $S_{\rm VIP}$ with $D\le 10$ Mpc. The modes of the distributions are almost unchanged. Even if the two skeletons are more similar, the skeleton $S_{\rm VIP}$ tends to oscillate around its true location (as reconstructed by $S_{\rm Par}$), making the length {\it L$_{\rm VIP}$} longer than {\it L$_{\rm Par}$}.

In conclusion, the close match between the two PDFs indicates that the skeleton reconstructed at 3$\sigma$ for the VIPERS-like catalogue is able to detect the most robust filaments (corresponding to a $5.5\sigma$ persistence threshold in a fully sampled dataset) and contains a small fraction of spurious filaments.

\section{Results}
\label{results}
\paragraph*{The filamentary structures of the Cosmic Web.}

\begin{figure*}
\includegraphics[width=\textwidth]{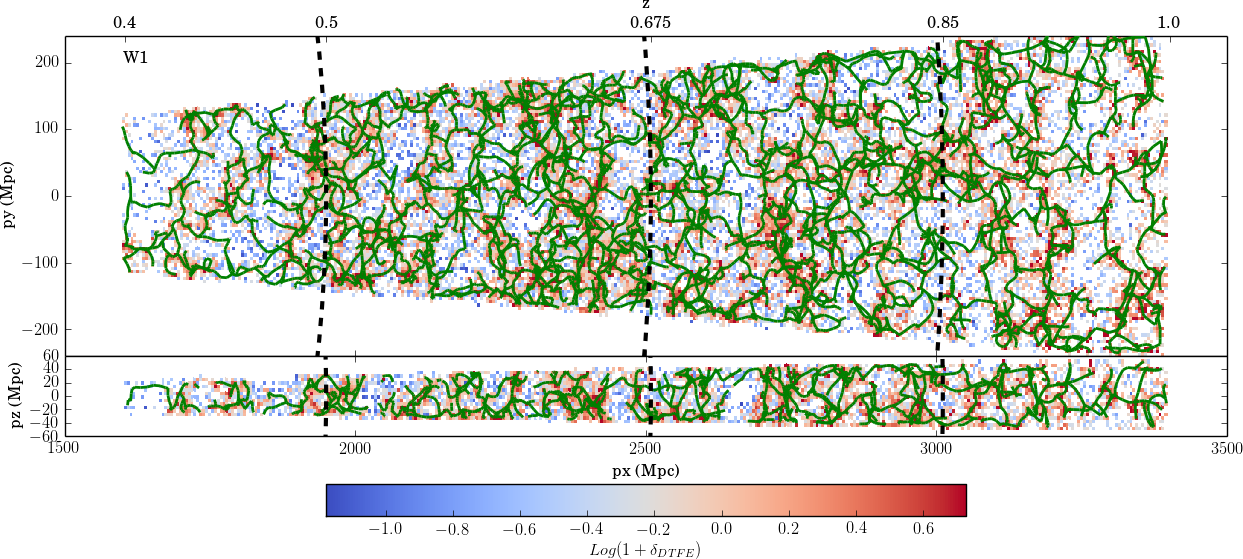}
\includegraphics[width=\textwidth]{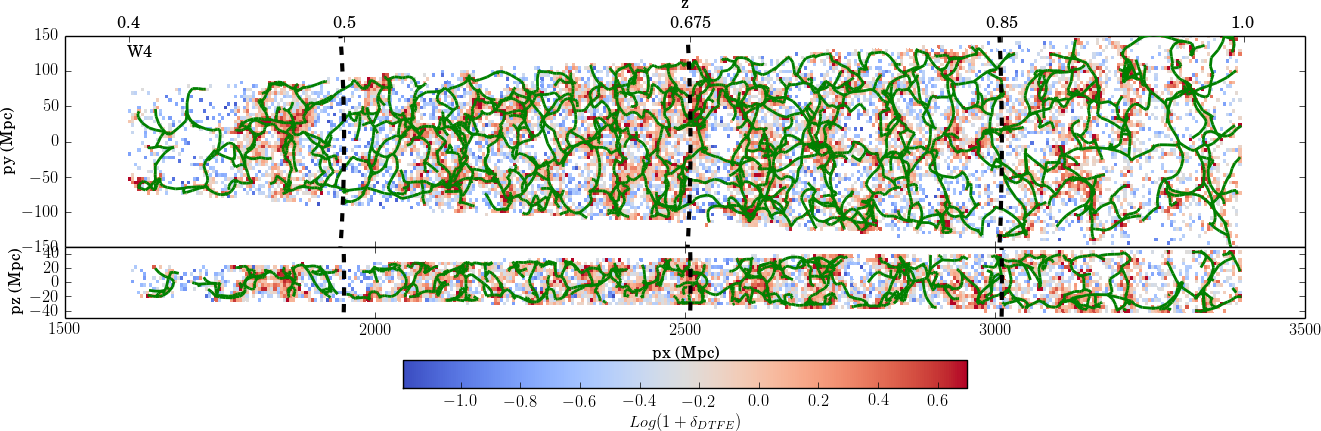}
\caption{Projected distribution of the filaments reconstructed with DisPerSE (in dark green) in the VIPERS W1 (top panel) and W4 (bottom panel) fields between $0.4 \le z\le 1$. The density contrast, $\log(1+\delta_{\rm DTFE})$, is averaged on cells of $5\times 5$ Mpc$^2$ and colour-coded as indicated (white for empty cells). \textit{Top rows}: projected distribution along the declination direction ($\Delta\delta=2$). \textit{Bottom rows}: projected distribution along the right ascension direction (in the central regions with $\Delta\alpha=2$). {\it 3D movies are available on the VIPERS website.}}
\label{W1W4_filaments}
\end{figure*}

We run  DisPerSE  on the VIPERS fields with a $3\sigma$ persistence threshold. Fig. \ref{W1W4_filaments} shows the filamentary network (green lines) for W1 (top panels) and W4 (bottom panels) fields, overplotted on a map of the density contrast $\delta={n_{\rm DTFE}}/{\overline{n}(z)}-1$, where $n_{\rm DTFE}$ is the local DTFE density estimate. Even in the 2D projections, we can see that filaments reveal the ridges of the 3D density field which, by construction, connect the density peaks between them via saddle points. The averaged length of the skeletons are similar with $L\sim$ 0.0013 and 0.0016 Mpc/Mpc$^3$ in W1 and W4 respectively. At low ($z \le 0.5$) and high redshift ($z \ge 0.85$), the number of filaments drops as a consequence of the lower sampling and only the most secure filaments are detected, as expected with DisPerSE. Thanks to the large contiguous volume probed by VIPERS, large voids, partly delineated by the filaments, are visible in both fields with radii as large as $R \sim 30$ Mpc (see \citealt{Micheletti2014} and the parallel paper by \citealt{Hawken2016} for void analysis in VIPERS).

\begin{figure*}
\includegraphics[width = \textwidth]{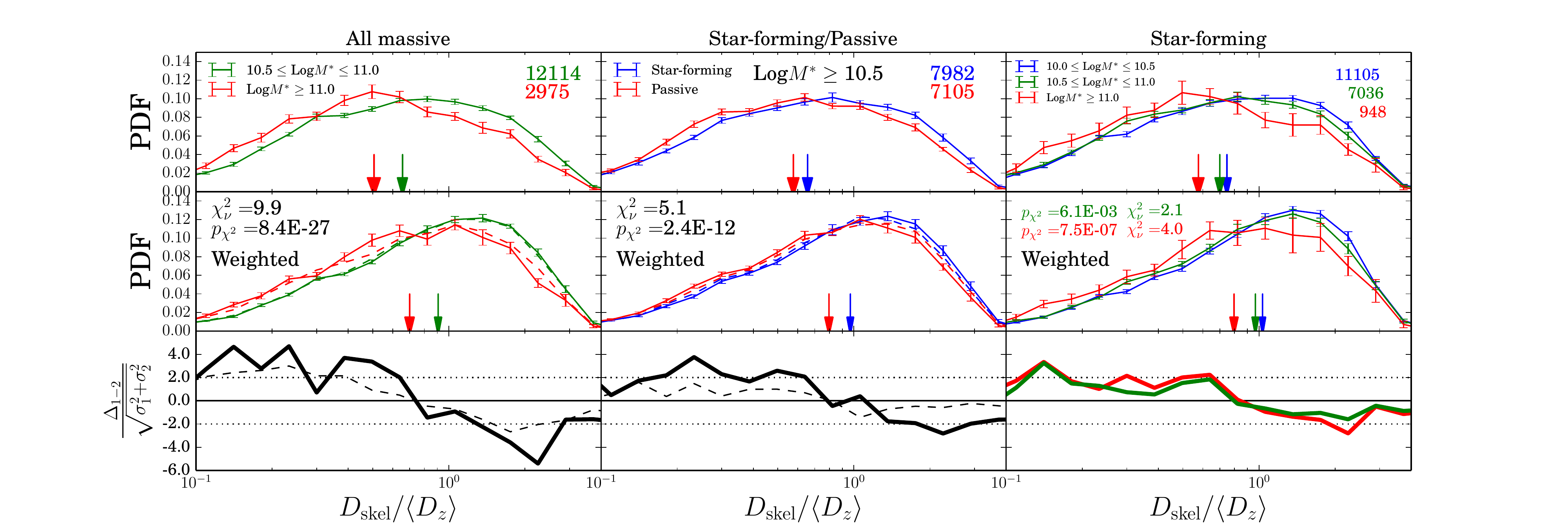}
\caption{Distributions of the unweighted (top row) and density-weighted (middle row) normalised distances ($D_{\rm skel} / \langle D_{z} \rangle$), for three selections: low \textit{vs} high mass galaxies (left column), quiescent \textit{vs} star-forming galaxies (middle column) and star-forming galaxies in three mass bins (right column). Vertical arrows indicate the medians of the PDFs and errorbars represent the dispersions computed with block-wise jackknife resampling. The PDFs after reshuffling of the samples (see text) are shown with dashed lines. The differences between the weighted distributions are shown in the bottom row. For star-forming galaxies (right panel) only the differences between high-intermediate (green) and high-low (red) mass bins are shown. The values resulting from the $\chi^2$ test of independence are reported in the middle panels.}
\label{dskelmass}
\end{figure*}

\paragraph*{Mass segregation inside filaments.}
We now investigate how different galaxy populations are distributed within this filamentary network in the redshift range $0.5 \le z \le 0.85$. We measure the distance of each galaxy to the nearest filament, $D_{\rm skel}$ (which, in the figure, we report normalised to $\langle D_{z} \rangle$ to take care of the variations of the mean inter-galaxy separation across the redshift range considered). The normalised PDFs of the distances, in the W1 and W4 fields combined, are shown in Fig. \ref{dskelmass} for three  selections: low \textit{vs} high mass galaxies, star-forming \textit{vs} quiescent galaxies, and star-forming galaxies in three mass bins. The errorbars are estimated with a block-wise ($1 \deg^2$) jackknife resampling. The first two samples are limited to $M^{\ast}\ge 10^{10.5}M_{\sun}$ to be complete in stellar mass for both quiescent and star-forming galaxies \citep[see][]{Davidzon2016} while a mass cut of $M^{\ast}\sim 10^{10}M_{\sun}$ is used when only star-forming galaxies are considered.

A trend between distance and stellar mass is observed (left column), with more massive galaxies being closer to filaments as indicated by the shift in the median values of the two PDFs (downward arrows). Passive galaxies are also found to be closer to filaments (middle column). While a large fraction ($47\%$) of our massive ($M^{\ast} \ge 10^{10.5} M_{\sun}$) galaxies are also passive, by looking at the star-forming population alone we observe a similar trend, albeit weaker, with the most massive star-forming galaxies being closer to filaments (right column). However, since we wish to evaluate the impact of the filaments on galaxy properties, we have to take into account the contribution of the nodes of the density field, usually related to galaxy groups and clusters, which are at the intersections of filaments and are known to be privileged regions where quenching is more efficient. This task is not easy, as there are partial overlaps between the local density and the CW environment \citep{AragonCalvo2010c}. As proposed by \citet{Gay2010}, to minimize the node contributions we weight each galaxy by the inverse of the density field smoothed using a Gaussian filter with $\sigma = 3$ Mpc. The weighted PDFs are shown in the middle row of Fig. \ref{dskelmass}. A shift in the medians of the PDFs to larger distances is observed but the trends remain. We also adopted alternative approaches by rejecting galaxies in high density regions ($\delta \ge 4$), located within groups, according to a parallel analysis (Iovino et al., in prep.), or by keeping only the filaments with a higher persistence threshold. They do not qualitatively change the results discussed in this section.

The significance of the observed trends is illustrated by the residuals between the weighted distributions expressed in units of $\sigma$ (bottom row of Fig. \ref{dskelmass}). The deviations exceed $2\sigma$ in most of the bins except for star-forming samples alone (due to shot noise in the most massive bin). We also quantify the differences with the $\chi^2$ test of independence and the probabilities of observing such a difference by chance which are negligible (listed in middle panels). This confirms the existence of a weak but statistically significant segregation effect inside the filaments and suggests that galaxy processing happens also during the drift of galaxies towards the nodes of the CW.

In Fig. \ref{dskelmass} we also look at how the mass-density relation is hidden in the observed mass segregation. We split the sample in local density bins and reshuffle the stellar masses between the galaxies in each bin. The mean PDF distributions for 10 random reshufflings are shown as dashed lines on the middle left panel. The PDFs for the low and high mass bins are close to the original ones, which shows that the mass segregation exists even after reshuffling the masses, if the mass-density relation is preserved. Therefore the mass segregation inside the filaments emerges naturally from the mass-density relation and the anisotropic distribution of the density in the CW.

A similar approach is adopted for the galaxy type segregation. We randomly attribute a galaxy type (passive/star-forming) to galaxies by conserving the type fraction observed in different stellar mass bins. The mean PDFs for 10 random reshufflings are shown as dashed lines (middle panel). In this case the segregation almost vanishes, with a difference between the two PDFs of less than $1 \sigma$.  The observed type segregation therefore does not arise just from the mass-type fraction relation but could have its origin in the dynamic of the large scale anisotropic structures of the CW.

\section{Discussion and conclusions}
\label{conclusions}
We reported the first characterization of large scale filamentary structures at $z \sim 0.7$, carried out in the cosmological volume probed by the VIPERS spectroscopic survey. The reconstruction is based on the DisPerSE code and the capability of VIPERS to capture such a CW's filamentary network is tested on simulations. We observe a small but significant trend for galaxies with different stellar masses and stellar activity to segregate near the filaments with the most massive and/or passive galaxies being closer to filaments. The signal persists even after down-weighting the contribution of nodes and high density regions.

The galaxy segregation quantified in this paper is a first step in support of a new paradigm in galaxy formation where large scale cosmic flows play a role in shaping galaxy properties. Beyond the observed anisotropy of the mass distribution (which follows naturally from the mass-density relation and the anisotropy of density in the CW), we expect that other physical parameters (\textit{e.g.} stellar activity controlled in part by gas accretion or morphology and size controlled in part by angular momentum) will be impacted by this large scale environment. Indeed, our results are fully consistent with the ingredients of the spin alignment theory presented in \cite{codis2015} which relies on these large-scale cosmic flows. The stellar activity segregation observed here can be interpreted inside this theory. Low mass or star-forming galaxies are preferentially located at the outskirts of filaments, a vorticity rich environment \citep{laigle2015}, where galaxies acquire both their angular momentum (leading to a spin parallel to the filaments) and their stellar mass essentially via smooth accretion \citep{welker2015}. The stellar-mass segregation observed here is the next step, where at higher mass, a transition is predicted in simulations, when more massive post-mergers drifting along filaments convert the orbital momentum of their progenitors into spin perpendicular to the axis of the filament, with increased efficiency for higher merger rate \citep{Dubois2014, welker2015}. The  most massive galaxies, dominated by the quiescent population, should preferentially  complete their stellar mass assembly in the core of filaments by merging.

We plan to extend the characterization of the CW in VIPERS in a future paper. It will be of interest to extend our analysis to other physical quantities such as star-formation rate  or  specific star-formation rate,  spin orientation and extent to different CW perspective:  the distance to nodes within the filaments  or, of particular interest, the distance from the saddle of the filaments, expected to be the region in the filament where galaxies have been the less affected by environmental effects. This will become within reach with the upcoming large and deep spectroscopic surveys such as PFS (Prime Focus Spectrograph). With its higher sampling and high-redshift extension ($z \sim 1.5-2$), PFS-deep survey will offer the opportunity to explore such dependencies near the peak of the cosmic star formation activity with an unprecedented accuracy.

Meanwhile, alternative approaches based on large multi-band photometric surveys (\textit{e.g.} COSMOS, J-PAS, \citealt{Benitez2014}) with accurate photometric redshifts allow us to analyze the projected 2D CW in narrow redshift slices. Preliminary results on the filamentary structures in the COSMOS field yield consistent results for the mass segregation inside the filaments (Laigle et al., in prep.).

\section*{Acknowledgments}
This work is based on observations collected at the European Southern Observatory, Cerro Paranal, Chile, using the Very Large Telescope under programmes 182.A-0886 and partly 070.A-9007. Also based on observations obtained with MegaPrime/MegaCam, a joint project of CFHT and CEA/DAPNIA, at the Canada-France-Hawaii Telescope (CFHT), which is operated by the National Research Council (NRC) of Canada, the Institut National des Sciences de l'Univers of the Centre National de la Recherche Scientifique (CNRS) of France, and the University of Hawaii. This work is based in part on data products produced at TERAPIX and the Canadian Astronomy Data Centre as part of the Canada-France-Hawaii Telescope Legacy Survey, a collaborative project of NRC and CNRS. The VIPERS web site is \href{http://www.vipers.inaf.it}{http://www.vipers.inaf.it}. This research is carried out within the framework of the Spin(e) collaboration (ANR-13-BS05-0005, \href{http://cosmicorigin.org}{http://cosmicorigin.org}). NM acknowledges the financial contributions by grants ASI/INAF I/023/12/0 and PRIN MIUR 2010-2011 \textquotedblleft The dark Universe and the cosmic evolution of baryons: from current surveys to Euclid\textquotedblright.

\bibliographystyle{mnras}
\bibliography{Malavasi4rasbib2}

\bsp

\label{lastpage}

\end{document}